\documentclass{ws-p9-75x6-50}
\newcommand{\Pom}{{I\!\!P}}
\newcommand{\R}{{I\!\!R}}
\newcommand{\xpom}{x_{\Pom}}

\newcommand{\as}{\alpha_s}

\begin{document}

\title{
A phenomenological study of BFKL evolution
\footnote{Invited talk given at the Hadron 13 Conference, Mumbai,
13-20 January 1999}}

\author{C. Royon}

\address{
CEA, DAPNIA, Service de Physique des Particules, \\
Centre d'Etudes de Saclay, France}


\maketitle

\abstracts{
The QCD dipole picture allows to build an unified
theoretical description -based on BFKL dynamics- of the total and
diffractive nucleon structure functions measured at HERA.
We use a four parameter fit to describe the 1994 H1
proton structure function $F_{2}$ data in the low $x$, moderate $Q^{2}$
range. The diffractive dissociation
processes are discussed within the same framework,
and a 6 parameter fit of the 1994 H1 diffractive structure 
function data is performed. \\
The BFKL dynamics can also be successfully tested at the $e^+e^-$
collider LEP and a
future high energy linear collider. The total $\gamma^*\gamma^*$
cross-section is calculated in the
Leading Order QCD dipole picture of BFKL  dynamics, and compared
with the one from 2-gluon exchange.
Next to Leading order corrections to the
BFKL evolution have been  determined phenomenologically,
and are found to  give very large
corrections to the BFKL cross-section,
leading to a reduced sensitivity
for observing BFKL effects.
The $Y$ dependence of the cross-section remains
a powerful tool to increase the ratio 
between the BFKL and the 2-gluon cross-sections and is more
sensitive to BFKL effects, even in the presence of large higher
order corrections. 
}

\section{Description of the proton structure function $F_2$ in the BFKL
framework}
To obtain the proton structure function $F_{2}$, we use the $k_{T}$
factorisation theorem, valid
for QCD at high energy (small $x$),
in order to factorise the $(\gamma~ g(k) \rightarrow q ~ \bar{q})$ cross
section and the
unintegrated gluon distribution of a proton containing the
physics of the BFKL pomeron \cite{Mueller,bfkl}.
The detailed calculations can be found in \cite{ourpap}.
\par
We finally obtain:
\begin{eqnarray}
\label{predF2}
F_2 \equiv F_T + F_L
= {\cal N} a^{1/2} e^{(\alpha_{\Pom} -1) \ln\frac{c}{x}} \frac{Q}{Q_0}
e^{- \frac{a}{2} \ln^2 \frac{Q}{Q_0}}
\end{eqnarray}
where $x$ and $Q^2$ are respectively the momentum fraction of the 
interacting quark and the exchanged energy squared,
$\alpha_{\Pom} -1 = \frac{4 \bar{\alpha} N_{C} \ln 2}{\pi}$, and
$a(x)=\left(\frac{\bar{\alpha} N_c}{\pi} 7 \zeta(3) \ln\frac{c}{x}\right)
^{-1}$.
The free parameters for the fit of the H1 data are ${\cal N}$,
the normalisation,
$\alpha_{\Pom}$, the pomeron intercept,
$Q_{0}$, and $c$.
\par
In order to test the accuracy of the $F_{2}$ parametrisation obtained in
formula
(1), a fit using the recently published data from the H1 experiment
\cite{H1} has been
performed \cite{ourpap}.
We have only used the points with $Q^{2} \leq 150 GeV^{2}$ to remain
in a reasonable domain of validity of the 
QCD dipole model. 
The $\chi^{2}$ is 88.7
for 130 points, and the values of the parameters are 
$Q_{0}=0.522 GeV$, ${\cal N}= 0.059,$ and $c=1.750,$ while
$\alpha_{\Pom}=0.282$.
Commenting on the parameters, let us note that  the effective coupling
constant
 extracted using (3) from $\alpha_{\Pom}$ is $\alpha =0.11$, close
to $\alpha (M_{Z})$ used in the H1 QCD fit. It is an acceptable value
for the 
small fixed value of the coupling constant required by the  BFKL
framework.
The running
of the coupling constant is not taken into account in the present BFKL
scheme. 
This could explain the rather low value of the effective $
\alpha _{\Pom}$ which is expected to be decreased by the next leading 
corrections.
The value of $Q_{0}$ corresponds
to a tranverse size of 0.4 fm which 
is in the correct range for a proton non-perturbative characteristic
scale. The value of ${\cal N}$ determines the amount of primordial
dipoles in 
the proton to be 
\begin{eqnarray}
\omega\left(1/2 \right) = {\cal N}\ \frac{128}{11\pi\ \alpha^2 N_c
e^2_f}\ 
\sqrt \frac{\pi}2 \simeq 7.55/e^2_f ,
\nonumber
\end{eqnarray}
The parameter $c$ sets the ``time'' scale for the formation of the
interacting 
dipoles. It defines the effective total rapidity interval which is $\log
(1/x) 
+ \log c,$ the constant being not predictible (but of order 1) at the
leading 
logarithmic
approximation.

\section{Diffractive structure functions}
In the dipole approach, two components contribute to the diffractive
structure function. First, the elastic component
corresponds to the elastic interaction of two dipole configurations.
It is expected to be
dominant in the finite $\beta$ region, i.e. for small relative masses 
of the diffractive system. 
Second, there is an inelastic component where the initial
photon dipole configuration is diffractively dissociated in multi-dipole states
by the target. This process is expected to be important at
small $\beta$ (large masses).
\par
The expressions for the elastic component is the following:
\begin{eqnarray}
F_T^{D(el)}=12\frac{N_ce^2 \alpha_s^4}{\pi}
\xpom^{-2\alpha_\Pom+1}\;a^3(\xpom)
\log^3\frac{Q}{2Q_0\sqrt{\beta}}\;e^{-a(\xpom)\log^2\frac{Q}{2Q_0\sqrt{\beta}}}\\
\times\beta(1\!-\!\beta)\;
\left[_2 F_1\left(-\frac{1}{2},\frac{3}{2};2;1-\beta\right)
\right]^2\ ,
\label{ftdqef}
\end{eqnarray}
\begin{eqnarray}
F_L^{D(el)}=16\frac{N_ce^2\as^4}{\pi}\;
\xpom^{-2\alpha_\Pom+1}\; a^3(\xpom)\log^2\frac{Q}{2Q_0\sqrt{\beta}}
\;e^{-a(\xpom)\log^2\frac{Q}{2Q_0\sqrt{\beta}}}\\
\times\beta^2\;
\left[_2 F_1\left(-\frac{1}{2},\frac{3}{2};1;1-\beta\right)\right]^2\ ,
\label{fldqef}
\end{eqnarray}
with $\alpha_{\Pom}$, and $a(\xpom)$ defined in the first section.
The inelastic component reads:
\begin{equation}
F_{T,L}^{D(in)}=2^9\sqrt{\frac{2}{\pi}} 
H_{T,L}\left(\frac{1}{2}\right)\frac{N_ce^2\as^5}{\pi^4}
\;\xpom^{-2\alpha_\Pom+1}a^3(\xpom)\frac{Q}{Q_0} 
e^{-\frac{a(\beta)}{2}\log^2\frac{Q}{4Q_0}}
 a^{\frac{1}{2}}(\beta)\beta^{1-\alpha_{Pom}}\ .
\label{fdinf}
\end{equation}
With this model, we are ready to write a full parametrization adequate for the 
description of the data.
The free parameters of the dipole model are $\alpha_\Pom$, which is related to
the fixed coupling constant $\as$ in the BFKL scheme at leading
order, $Q_0$, corresponding to a non-perturbative scale
for the proton, and the three normalizations $N_T^{(el)}$, $N_L^{(el)}$, $N^{(in)}$.
As it is now well-known, a secondary
trajectory based on reggeon exchange is added in order to take into account the
large-mass and small rapidity gap domain.
Reggeon exchange can here be simply parametrized in the following way:
$F_2^{D(R)} (x_P, \beta, Q^2) =
 f^\R (x_\Pom) F_2^\R (\beta, Q^2)$ ,
where the reggeon flux $f^\R (x_\Pom)$ is assumed to follow a Regge behaviour with a linear
trajectory $\alpha_{\R}(t)=\alpha_{\R}(0)+\alpha^\prime_{\R}\;t\;$:
$f^{\R}(x_{\Pom})= \int^{t_{min}}_{t_{cut}} dt\;\frac{e^{B^{\R}\; t}}
{x_{\Pom}^{2 \alpha_{\R}(t) -1}}$,
where $|t_{min}|$ is the minimum kinematically allowed value of $|t|$
and $t_{cut}=-1$ GeV$^2$ is the limit of the measurement. The values
of $B^{\R}$ and $\alpha^\prime_{\R}$ are fixed with data from hadron-hadron
collisions \cite{F2DH194}. The reggeon structure function is assumed 
to be the pion structure function \cite{grv}. The free parameters for 
this component are the reggeon normalisation $N^\R$ and 
exponent $\alpha_{\R}$.

A fit to the recently published H1 \cite{F2DH194} and ZEUS
\cite{zeus} diffractive structure 
function data is performed separately. The result of the H1 fit is
shown in Figure 1. The fit to the H1 data leads
to a very good $\chi^2$ (1.17 per degree of freedom with statistical errors
only). The parameters and the features of the fits are given 
in detail in Ref. \cite{ourf2d}, and we will comment here only on the main 
parameters. The value of $\alpha_\R$ ($\alpha_\R$=0.68) is consistent 
with the usual values found for secondary reggeon contributions if 
interference effects
are not taken into account \cite{F2DH194}. The value of $\alpha_\Pom$
($\alpha_\Pom$=1.40)
is found to be consistent with the expected intercept for a hard BFKL pomeron
\cite{bfkl}. This intercept is higher than the value obtained from the
fit to the structure function $F_2$.  $Q_0$ 
($Q_0$=0.43) is a typical non 
perturbative scale for the
proton and very close to the value obtained in the proton structure function
fit. It should be noted that the scale $Q_0$ appears in a quite non trivial
way as the virtuality in the inelastic component ($Q/4Q_0$), and in the
elastic one ($Q/2 \sqrt{\beta} Q_0$). 
\par
The fit to the ZEUS data leads to a worse $\chi^2$ ($\chi^2 /dof=
1.95$ with statistical errors only). 
In order to investigate the origin of these differences, a direct comparison
between ZEUS and H1 data has been performed \cite{ourf2d}. The H1 data have
been interpolated to the ZEUS closest bins in $\beta$ and $Q^{2}$ using the
dipole model fit. This interpolation is weakly sensitive to the model used
as the interpolation in the kinematical variables is very small. It was 
checked that the use of the model 
by Bartels et al. \cite{bartels} gives a similar result.
The striking feature is that the main difference
between the H1 and ZEUS fits comes from the region where the data are 
different, showing that this region influences the global fit.
\par
Let us now comment about some of the main features of the fit. Some additional
details and discussions can be found in \cite{ourf2d}.
First, the effective
intercept of the proton diffractive structure function in the QCD dipole picture
is clearly not consistent with the soft pomeron value
(1.08), but much lower than the bare pomeron intercept obtained in the 
fit ($\alpha_{\Pom}=1.40$). This can be explained by the large logarithmic
corrections induced by the $a^3(x_\Pom)$ term, proportional to
$\log^3(1/x_\Pom)$, present in both diffractive components (see formulae 
\ref{ftdqef}, \ref{fldqef}, \ref{fdinf}). The effect of this logarithmic
term induces also an $\xpom$ dependence of the intercept. Moreover, 
the $\xpom$ dependence of the intercept is different 
between the elastic and the inelastic components. This induces a breaking of
factorisation directly for the diffractive components of this model, which 
comes in addition to the known factorisation breaking due to secondary 
trajectories. 

One interesting feature of the diffractive proton structure functions was 
the $Q^2$ dependence at fixed $x_{\Pom}$ as a function of $\beta$ as was
pointed out experimentally by the H1 collaboration \cite{F2DH194} 
and
confirmed at lower $Q^2$.
In the QCD dipole model, this experimental feature is described by a non
trivial interplay between the two diffractive components. In Figure 2,
the dipole fit is compared with the H1 result showing the contribution
of each component: at small $\beta$, the inelastic component dominates and
varies quasi linearly in $\log Q^2$, and at high $\beta$, this component
is depressed similarly to the total structure function, but is 
progressively substituted
by the elastic component.

One important result of the dipole model is also the fact that the longitudinal 
elastic component is
found to be high at high $\beta$, which leads to high values of the
ratio $R$ of the longitudinal to the transverse components at high values 
of $\beta$. We obtain that the $R$ ratio remains small ($\sim 0.2$) in almost the full
kinematical plane except notably at high $\beta$ where it may reach high values
such as 2. Note that this value is in the range of the measured $R$
ratio with vector meson production. A measurement of $R$
in diffraction would thus be of great interest and would be a good test
of the model. It is instructive to notice that another model of diffraction based on
selecting $q \bar{q}$ and $q \bar{q} g$ components of the photon
\cite{bartels} also leads to a large contribution of 
the longitudinal $q \bar{q}$ contribution at high $\beta$.

\section{$\gamma^* \gamma^*$ total cross-section}
Here, we want to calculate the total $\gamma^*\gamma^*$ cross-section derived in the 
Leading Order QCD dipole picture of BFKL dynamics. This could be a good test
of the BFKL equation which an be performed at e$^+$-e$^-$ colliders (LEP or
linear collider LC). The advantage of this process compared to the ones
discussed in the previous sections is that it is a purely perturbative process.

In this study, we compare the 2-gluon and the BFKL cross-sections. Defining 
$y_1$ (resp. $y_2$), and $Q_1^2$ (resp. $Q_2^2$) to be the rapidities and
the squared transfered energies for both virtual photons, one gets
\begin{eqnarray}
\label{eeBFKLa}
&&\hspace{-1.5cm}d\sigma_{e^+e^-}
 (Q_1^2,Q_2^2;y_1,y_2) = \frac{4}{9} \left(\frac{\alpha_{e.m}^2}{16}\right)^2  \, 
 \alpha_s^2 \, \pi^2 \sqrt{\pi} \,
\frac{d Q_1^2}{Q_1^2} \frac{d Q_2^2}{Q_2^2} \frac{d y_1}{y_1} 
\frac{d y_2}{y_2}
\frac{1}{Q_1 \, Q_2}
\, \nonumber \\
\frac{e^{\displaystyle \frac{4\alpha_s N_c}{\pi} Y \ln 2 }}{\sqrt{\frac{14
\alpha_s N_c}{\pi} Y \zeta(3)}} \, 
&& \times \, e^{\displaystyle -\frac{\ln^2 \frac{Q_1^2}{Q_2^2}}{\frac{56 
\alpha_s N_c}{\pi} Y \zeta(3)}}
 \left[2 l_1  +
9 t_1 \right] \, \left[2 l_2  +
9 t_2 \right] \,,
\end{eqnarray}
for the BFKL-LO cross-section, where
$t_1 = \frac{1 + (1-y_1)^2}{2}, \quad l_1=1-y_1$, and an analogous definition
for $t_2$ and $l_2$, 
and  
$Y=\ln \frac{s y_1 y_2 }{\sqrt{Q^2_1 Q^2_2}}$.
The 2-gluon cross-section has been calculated exactly within the high energy
approximation (NNNLO calculation) and reads
\begin{eqnarray}
\label{2g}
&~& d\sigma_{e^+e^-}
 (Q_1^2,Q_2^2;y_1,y_2) = \frac{d Q_1^2}{Q_1^2} \frac{d Q_2^2}{Q_2^2} 
\frac{d y_1}{y_1} \frac{d y_2}{y_2} 
 \, \frac{64 (\alpha_{e.m}^2 \alpha_s)^2}{243 \pi^3} \frac{1}{Q_1^2} 
\nonumber \\
&~& \left[ t_1 t_2 \ln^3 \frac{Q_1^2}{Q_2^2} + \left( 7 t_1 t_2 
+ 3 t_1 l_2 + 3 t_2 l_1 \right) \ln^2 \frac{Q_1^2}{Q_2^2} \right. \nonumber
\\ &~&  \left.
+ \left( ( \frac{119}{2} - 2 \pi^2 ) t_1 t_2
+ 5 (t_1 l_2 +t_2 l_1) + 6 l_1 l_2 \right) \ln \frac{Q_1^2}{Q_2^2} \right.
\nonumber
\\ &~& \left.
+ \left( \frac{1063}{9} - \frac{14}{3} \pi^2 \right) t_1 t_2
+ (46 - 2 \pi^2) (t_1 l_2 + t_2 l_1) - 4 l_1 l_2 \right] \,.
\end{eqnarray} 

Rsults based on the calculations developed above 
will be given for 
 LEP (190 GeV centre-of-mass energy) and  a future Linear Collider 
(500 - 1000 GeV centre-of-mass energy).
$\gamma^* \gamma^*$ interactions are selected at $e^+e^-$ colliders by detecting the 
scattered electrons, which leave the beampipe, in forward
calorimeters. Presently at LEP these detectors can measure electrons 
with an angle $\theta_{tag}$ down 
to approximately 30 mrad. For the LC it has been argued~\cite{brl}
that angles as low as 20 mrad should be reached. Presently
angles down to 40 mrad are foreseen to be instrumented for a generic
detector at the LC. 

Let us first specify the
region of validity for the parameters controlling
the basic assumptions made in the previous chapter. The main constraints
are required by the validity of the perturbative calculations.
The ``perturbative'' constraints are imposed by considering only photon 
virtualities $Q^{2}_{1}$, 
$Q^{2}_{2}$ high enough so that the scale $\mu^{2}$ in $\alpha_S$ 
is greater than 3 GeV$^{2}$. $\mu^2$ is defined 
using the Brodsky Lepage Mackenzie (BLM) scheme \cite{blm}, 
$\mu^2=\exp(- \frac{5}{3}) \sqrt{Q_1^2 Q_2^2}$ \cite{blm}.
In this case 
$\alpha_S$ remains always 
small enough such that the perturbative calculation is valid. 
In order that gluon contributions dominates the QED one, $Y$ 
is required to stay
larger than $\ln(\kappa)$ with $\kappa = 100.$ (see Ref. \cite{blm} for 
discussion).
Furthermore, in order to suppress DGLAP evolution, while maintaining
BFKL evolution  will constrain  $0.5 < Q_1^2/ Q_2^2 < 2$
for all nominal calculations.  The comparison between the
DGLAP-DLL and the 2-gluon cross-section in the LO approximation shows that
 both cross-sections are similar when
$Q_1$ and $Q_2$ are not too different ($0.5 < Q_1^2/ Q_2^2 < 2$), so
precisely in the kinematical domain where the BFKL cross-section is 
expected to dominate. However,
when $Q_1^2/Q_2^2$ is further away from one, the LO 2-gluon cross-section 
is lower than the DGLAP one, especially at large $Y$. This suggests that
the 2-gluon cross-section could be a good approximation of the
DGLAP one if both are calculated at NNNLO and restricted
to the region where 
$Q_1^2/Q_2^2$ is close to one. In this paper we will use the
exact NNNLO 2-gluon cross-section in the following to evaluate the 
effect of the  non-BFKL background,
 since the 2-gluon term appears to constitute 
 the dominant part of the DGLAP cross-section in the 
region $0.5 < Q_1^2/Q_2^2 < 2$. 

We will not discuss here all the phenomenological results, and some detail can
be found in \cite{gamma}, as well as the detailed calculations.
We first study the effect of the tagged electron energy and angle.
We first study the effect of increasing the LC detector acceptance 
for electrons scattered under small angles and the ratio of the 2-gluon 
and the BFKL-LO cross-sections increase by more a factor 3 if the tagging 
angle varies between 40 and 20 mrad. The effect of lowering the tagging energy
is smaller. An important issue on the BFKL cross-section is the importance
of the NLO corrections and we adopt a phenomenological approach to estimate the effects
of higher orders. First, at Leading Order, the rapidity $Y$ is not 
uniquely defined, and we can add an additive constant to $Y$. This corresponds
to the $c$ parameter we discussed in the first section.
A second effect of NLO corrections is to lower the 
value of the so called Lipatov exponent in formula
\ref{eeBFKLa}. In the same $F_2$ fit described in section 1, the value of the 
Lipatov exponent $\alpha_\Pom$:
was fitted and found to be 1.282, which gives an effective value of $\alpha_s$
of about 0.11. The same idea can be applied phenomenologically for the $\gamma^* 
\gamma^* $ cross-section. For this purpose, we modify the scale in $\alpha_S$ so that
the effective value of $\alpha_S$ for $Q_1^2=Q_2^2=25$ GeV$^2$ is about
$0.11$. 
Finally, the results of
the BFKL and 2-gluon cross-sections are given in Table \ref{1.fin} if we
assume both effects. 
The ratio
BFKL to 2-gluon cross-sections is reduced to 2.3 if both effects are
taken into account together. In the same table, we also give these effects for LEP with the 
nominal selection and at the 
LC with a detector with increased angular acceptance. 
The column labelled 'LEP*' gives the results for the kinematic cuts 
used by the L3-collaboration who have recently  presented 
preliminary results \cite{l3}. The cuts are 
$E_{tag}$ = 30 GeV and $\theta_{tag}>30$ mrad and $\mu^2> 2$ GeV$^2$.
For this selected region the difference between NLO-BFKL and 2-gluon 
cross-section is only a factor of 2.4. A cut on $Q^2_1/Q^2_2$,
as done for the other calculations in this paper,  would help 
to allow a more precise determination of the 2-gluon 'background'.
Another idea to establish the BFKL effects in data is to study the 
energy or $Y$ dependence of the cross-sections, rather than the comparison with 
total cross-sections itself. To illustrate this point,
we calculated the BFKL-NLO and the 2-gluon cross-sections, as well as
their ratio, for given cuts on rapidity $Y$ (see table 
\ref{fin.fin}). We note that we can reach
up to a factor 5 difference ($Y \ge 8.5$) keeping a cross-section
measurable at LC. The cut $Y \ge$ 9. would give a cross-section hardly
measurable at LC, even with the high luminosity possible at this
collider. Cuts on $Y$ will be hardly feasible at LEP because of the
low cross-sections obtained already without any cuts on $Y$.

\begin{table}
 \begin{center}
\begin{tabular}{|c||c||c|c|c||c|} \hline
   & BFKL$_{LO}$     & BFKL$_{NLO}$ & 2-gluon     & ratio \\ \hline \hline
 LEP        & 0.57   & 3.1E-2       & 1.35E-2  & 2.3     \\
 LEP*        & 3.9   &  0.18       & 6.8E-2  & 2.6     \\
 LC 40 mrad & 6.2E-2 & 6.2E-3       & 2.64E-3  & 2.3     \\
 LC 20 mrad & 3.3    & 0.11         & 3.97E-2  & 2.8     \\ \hline
\end{tabular}
\caption{Final cross-sections (pb), for selections described in the text.}
\label{1.fin}
 \end{center}
\end{table}

\begin{table}
 \begin{center}
\begin{tabular}{|c||c||c|c|c||c|} \hline
$Y$ cut   & BFKL$_{NLO}$     &  2-gluon     & ratio \\ \hline
\hline
 no cut & 1.1E-2 & 3.97E-2 & 2.8 \\
 $Y \ge $ 6. & 5.34E-2 & 1.63E-2 & 3.3 \\
 $Y \ge $ 7. & 2.54E-2 & 6.58E-3 & 3.9 \\
 $Y \ge $ 8. & 6.65E-3 & 1.43E-3 & 4.7 \\
 $Y \ge $ 8.5 & 1.67E-3 & 3.25E-4 & 5.1 \\
 $Y \ge$ 9. & 5.36E-5 & 9.25E-6 & 5.8 \\ \hline
\end{tabular}
\caption{Final cross-sections (pb), for selections described in the
text, after different cuts on $Y$}
\label{fin.fin}
 \end{center}
\end{table}

\section{Conclusion}
Finally
the colour dipole model formalism calls for a unified description of the 
diffractive and  total
deep-inelastic scattering events.
We showed that within the precision of the current data, there are quite a few 
indications (similar scale $Q_0,$ softening of the hard Pomeron by logarithmic 
factors 
in diffraction,etc...) of such a common theoretical ground. 
However, further tests of the model are deserving. The first one would be a 
confrontation
of the predicted $R$ ratio with the data if available: indeed, the various models
should predict quite different contributions from the two polarization states
of the photon. Other useful tests concern the
final states. For instance one can compute the 
predictions for diffractive vector meson production and confront them to the 
recent data. Such tests might help distinguish between
the few different models for hard diffraction which are able 
to describe the data. 

The other topic discussed here is the difference between the 2-gluon and BFKL
$\gamma^* \gamma^*$ cross-sections 
both at LEP and LC. The LO BFKL cross-section is much larger than the 2-gluon cross-section.
Unfortunately, the higher order  corrections 
of the BFKL equation (which we estimated phenomenologically)
are large, and the 2-gluon and BFKL-NLO cross-sections ratios are 
reduced to a factor two to four.
The $Y$ dependence of the cross-section remains
a powerful tool to increase this ratio and is more 
sensitive to BFKL effects, even in the presence of large higher 
order corrections. Further more, the higher order corrections to the BFKL
equations were treated here only phenomenologically, and we noticed that
even a small change on the BFKL pomeron intercept implies large changes on
the cross-sections. The uncertainty on the BFKL cross-section after
higher order corrections is thus quite large. 
We thus think that the measurement performed at LEP
or at LC should be compared to the precise calculation of the 2-gluon
cross-section after the kinematical cuts described in this paper, and the
difference can be interpreted as BFKL effects. A fit of these
cross-sections will then be a way to determine the BFKL pomeron intercept
after higher order corrections.

\begin{figure}
 \begin{center}
\epsfxsize=35pc
\epsfbox{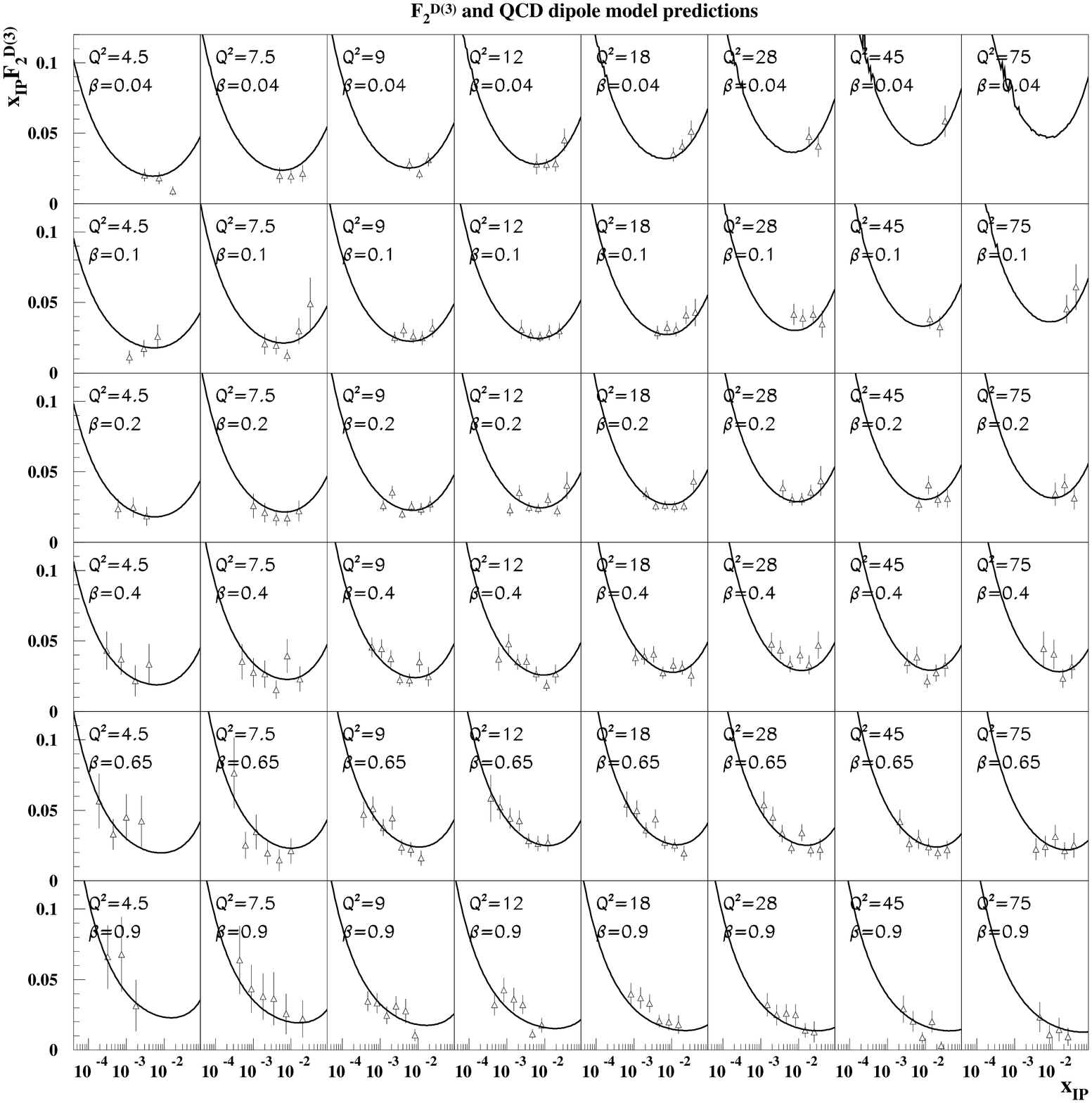} 
\end{center}
 \caption{Result of the $F_2^{D}$ fit to the H1 data. 
The data are displayed by
triangles (with statistical and systematic errors added in quadrature) as a function
of $x_\Pom$ in $\beta$ and $Q^2$ bins. The fit has been performed with
statistical errors only and is displayed in full line.}
 \label{theta_lc}
\end{figure}

\begin{figure}
 \begin{center}
\epsfxsize=20pc
\epsfbox{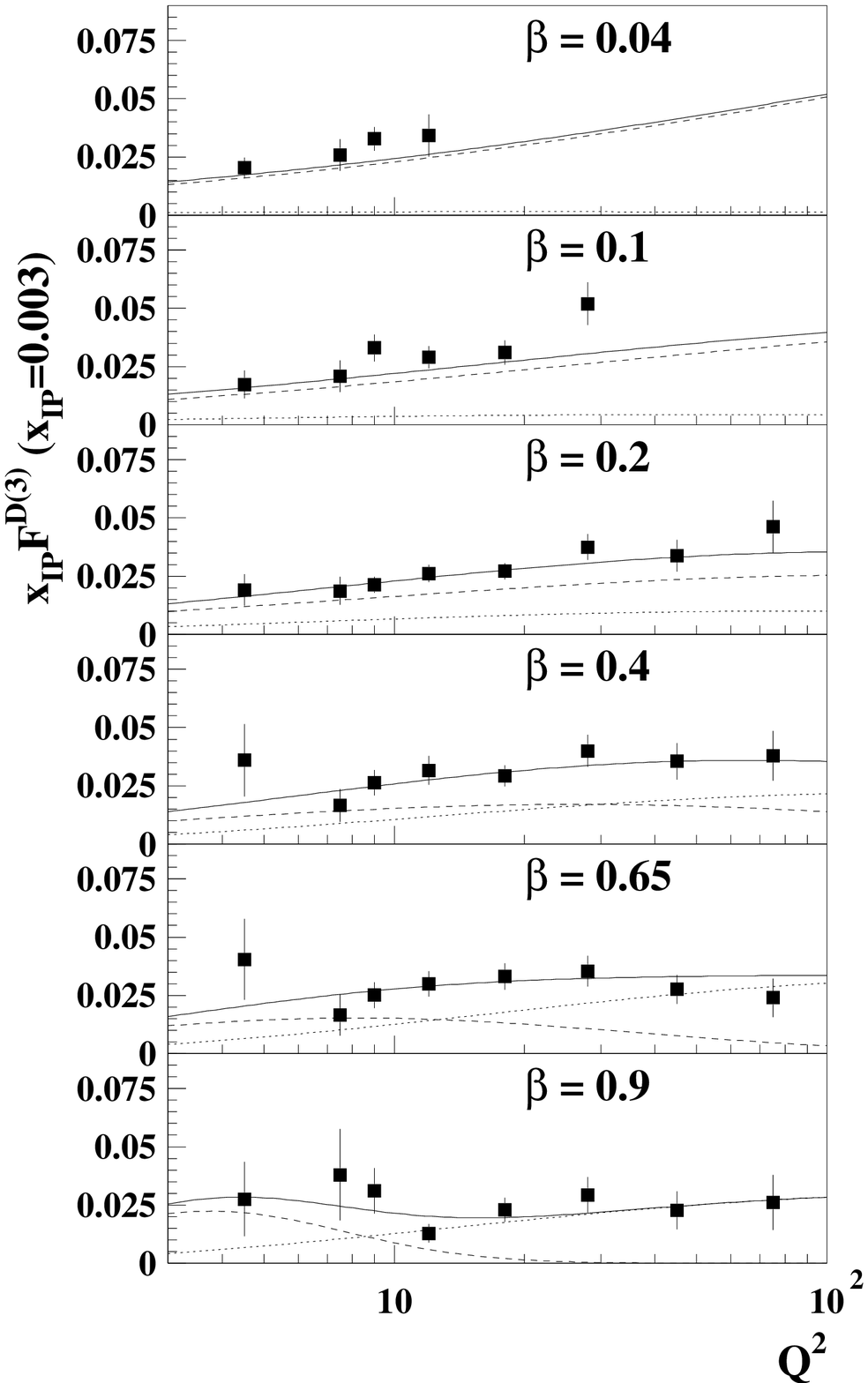} 
\end{center}
 \caption{Scaling violations. The dependence of $\xpom F_2^D$ on $Q^2$ for different values of
$\beta$ at fixed $\xpom$ ($3.10^{-3}$) is shown together with 
the dipole model fit. Dotted line: elastic component, dashed line:
inelastic component, full line: total.}
\end{figure}

\section{Acknowledgements}
The results described in the present contribution come from a fruitful
collaboration with A.Bialas, M.Boonekamp, S.Munier, H.Navelet, R.Peschanski, 
A.De Roeck, S.Wallon. I also thank the organisers for financial support.


\begin{thebibliography}{99}

\bibitem{Mueller}
A. H. Mueller, {\it Nucl. Phys.} {\bf B415} (1994) 373,
A. H. Mueller, B. Patel, {\it Nucl. Phys.} {\bf B425} (1994) 471,
A. H. Mueller, {\it Nucl. Phys.} {\bf B437} (1995) 107.

\bibitem{bfkl}
V. S. Fadin, E.A. Kuraev and L.N. Lipatov, {\it Phys. Lett.} {\bf B60} (1975) 50;\\
I. I. Balitsky and L.N. Lipatov, {\it Sov. J. Nucl. Phys.} {\bf 28} (1978) 822.

\bibitem{ourpap}
H. Navelet, R. Peschanski, Ch. Royon, S. Wallon, {\it Phys. Lett.} 
{\bf B385} (1996) 357, H.Navelet, R.Peschanski, Ch.Royon, {\it Phys.Lett.}, {\bf B366},
(1996) 329.

\bibitem{H1} H1 coll., {\it Nucl.Phys.} {\it B470} (1996) 3

\bibitem{F2DH194} 
C. Adloff et al., H1 coll., {\it Z.Phys.} {\bf C76} (1997) 613.

\bibitem{grv} 
M.Gl\"uck, E.Reya, A.Vogt, {\it Z.Phys.} {\bf C53} (1992) 651.
\bibitem{zeus}
J. Breitweg, ZEUS coll., {\it Eur. Phys. J.} {\bf C1} (1998) 81.

\bibitem{ourf2d}
A. Bialas, R. Peschanski, Ch. Royon, {\it Phys. Rev. }{\bf D57} (1998) 6899,
S.Munier, R.Peschanski, Ch.Royon, {\it Nucl.Phys.}
{\bf B534} (1998) 297.

\bibitem{bartels}
J. Bartels, J. Ellis, H. Kowalski, M. W\"usthoff, {\it Eur.Phys.J.}
{\bf C7} (1999) 443, 
J.Bartels, C.Royon, preprint DESY 98-152, {\tt hep-ph/9809344}.

\bibitem{brl} J.~Bartels, A.~De Roeck and H.~Lotter, 
Phys. Lett. {\bf B389} (1996) 742-748, J.~Bartels, A.~De Roeck and H.~Lotter,
C.~Ewerz, DESY preprint 97-123E, hep-ph/9710500.

\bibitem{blm} S.J.~Brodsky, F.~Hautmann and D.E.~Soper, 
Phys. Rev. Lett. {\bf 78} (1997) 803-806, Phys. Rev. {\bf D56} (1997) 6957-6979,
G.P.~Lepage, P.B.~Mackenzie, Phys. Rev. {\bf D48} (1993) 2250,
S.J.~Brodsky, G.P.~Lepage, P.B.~Mackenzie, Phys. Rev. {\bf D28} (1983) 228.

\bibitem{gamma}
M.Boonekamp, A. De Roeck, Ch. Royon, S.Wallon, preprint DAPNIA-SPP 99-01,
{\tt hep-ph/9812523}.

\bibitem{l3} Chu Lin, contributed talk at HADRON98, 
Stara Lesna, September 1998;\\
M.~Kienzle, contributed talk at the 2-photon workshop, Lund, September 
1998.

\end{thebibliography}
\end{document}